\documentclass[12pt]{article}
\usepackage{graphicx}
\usepackage{cite}
\usepackage{comment} 
\usepackage{mathtools}	
\usepackage{tikz} 
\newcommand*\circled[1]{\tikz[baseline=(char.base)]{\node[shape=circle,draw,inner sep=10pt, label={center:#1}] (char) {};}}

\newcommand\pubnumber{CP3-16-55}
\newcommand\pubdate{\today}

\def\institute{Centre for Cosmology, Particle Physics and Phenomenology (CP3), \\ Universit\'e Catholique de Louvain, Chemin du Cyclotron 2, B-1348 Louvain-la-Neuve, Belgium}

\def\Title#1{\begin{center} {\Large #1 } \end{center}}
\def\Author#1{\begin{center}{ \sc #1} \end{center}}
\def\Address#1{\begin{center}{ \it #1} \end{center}}

\newcommand\pubblock{\rightline{\begin{tabular}{l} \pubnumber\\
         \pubdate  \end{tabular}}}
\newenvironment{Abstract}{\begin{quotation}  }{\end{quotation}}
\newenvironment{Presented}{\begin{quotation} \begin{center} 
             PRESENTED AT\end{center}\bigskip 
      \begin{center}\begin{large}}{\end{large}\end{center} \end{quotation}}
\def\Acknowledgements{\bigskip  \bigskip \begin{center} \begin{large}
             \bf ACKNOWLEDGEMENTS \end{large}\end{center}}

\def\beq{\begin{equation}}
\def\eeq#1{\label{#1}\end{equation}}
\def\eeqn{\end{equation}}

\def\beqa{\begin{eqnarray}}
\def\eeqa#1{\label{#1}\end{eqnarray}}
\def\eeqan{\end{eqnarray}}

\let\bar=\overbar

\def\Dslash{\not{\hbox{\kern-4pt $D$}}}
\def\dslash{\not{\hbox{\kern-2pt $\del$}}}

\def\msb{{\bar{\ssstyle M \kern -1pt S}}}

\begin{document}
\begin{titlepage}
\pubblock
\def\thefootnote{\fnsymbol{footnote}}
\setcounter{footnote}{0}

\vfill
\Title{EW corrections to $t \bar t$ distributions: the photon PDF effect}
\vfill
\Author{ Ioannis Tsinikos\footnote{email: ioannis.tsinikos@uclouvain.be}}
\Address{\institute}
\vfill
\begin{Abstract}
In this proceeding we present the impact of the EW corrections on $t \bar t$ distributions focusing on the effect of the photon PDF. We use the {\sc NNPDF2.3QED} and {\sc CT14QED} PDF sets, which include the photon PDF. We discuss a detailed comparison between the results obtained from these two PDF sets at NLO QCD+EW accuracy for 13 TeV, focusing on the top-quark $p_T$ and the top-quark pair rapidity. We point out the differences between these two PDF sets and we show that the rapidity of the top-quark pair can be used in order to constrain the photon PDF predicted by the {\sc NNPDF2.3QED} set.
\end{Abstract}
\vfill
\begin{Presented}
$9^{th}$ International Workshop on Top Quark Physics\\
Olomouc, Czech Republic,  September 19--23, 2016
\end{Presented}
\vfill
\end{titlepage}
\def\thefootnote{\arabic{footnote}}
\setcounter{footnote}{0}

\section{Introduction}
\label{sec:intro}

The material discussed here is part of the project presented in \cite{Pagani:2016caq}. The main motivation for this work was the reported tension between theory and experiment at the high top-quark $p_T$ region at 8 TeV \cite{Khachatryan:2015oqa,Aad:2015mbv}. The QCD corrections beyond the NLO \cite{Czakon:2013tha,Czakon:2013goa,Czakon:2015owf,Czakon:2016dgf,Moch:2008qy,Kidonakis:2009ev,Czakon:2009zw,Ahrens:2010zv,Beneke:2010da,Kidonakis:2010dk,Ahrens:2011mw,Ahrens:2011px,Beneke:2011mq,Cacciari:2011hy,Pecjak:2016nee} to $t \bar t$ observables decrease the theoretical uncertainties. On top of that the experimental uncertainties will also decrease at the new run of LHC at 13 TeV and furthermore the $t\bar t$ process enters many LHC analyses as a signal or as a dominant background. These facts point to the necessity of the inclusion of NLO EW corrections to $t\bar t$ production. There is a vast literature in the subject \cite{Beenakker:1993yr,Kuhn:2006vh,Kuhn:2013zoa,Bernreuther:2005is,Campbell:2016dks,Hollik:2011ps,Kuhn:2011ri,Manohar:2012rs,Bernreuther:2012sx,Bernreuther:2010ny,Denner:2016jyo}, including also the photon-induced contributions at LO \cite{Hollik:2007sw}. These contributions are important because they are positive and therefore expected to possibly compensate part of the negative Sudakov suppression from the virtual EW corrections. For this reason there are many PDF sets developed including also the photon PDF. These are the {\sc MRST2004QED} \cite{Martin:2004dh}, the {\sc NNPDF2.3QED} \cite{Ball:2013hta}, the {\sc APFEL\_NN2.3QED} \cite{Bertone:2013vaa,Bertone:2015lqa}, the {\sc CT14QED} \cite{Schmidt:2015zda} and the more recent {\sc NNPDF3.0QED} \cite{Ball:2014uwa} and {\sc LUXqed} \cite{Manohar:2016nzj}.

\section{Different PDF sets}

\begin{figure}[htb]
\centering
\includegraphics[height=2.5in]{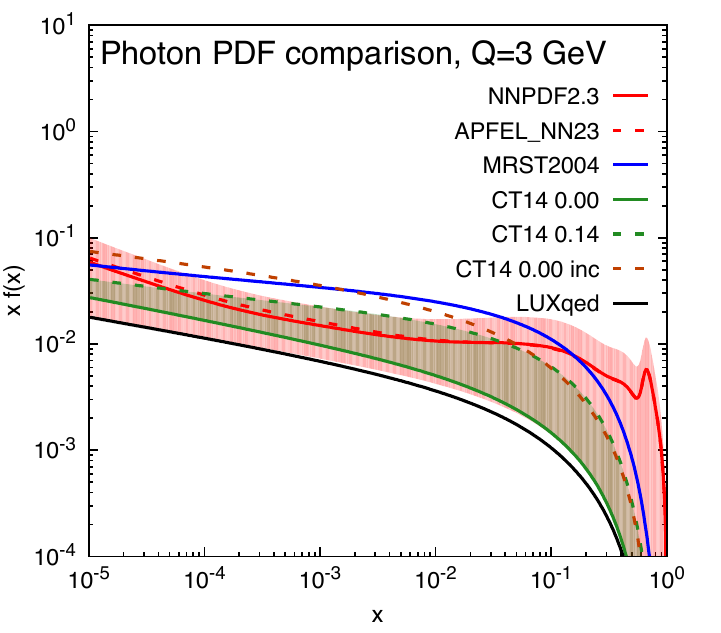}
\includegraphics[height=2.5in]{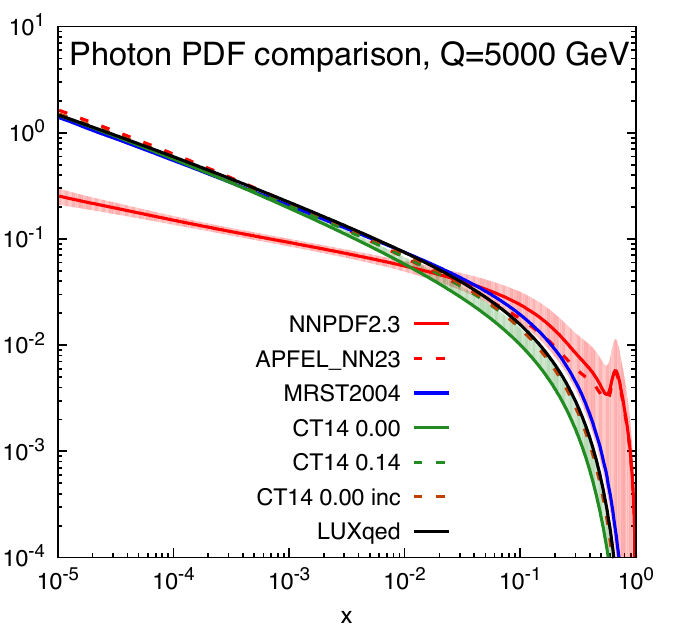}
\caption{Comparison of the photon PDF for different PDF sets at $Q=3$ GeV and $Q=5$ TeV.}
\label{fig:PDFs}
\end{figure}

All the aforementioned PDF sets include the LO QED evolution for the photon PDF, but the DGLAP QCD+QED evolution is realised differently. For this reason their comparison in fig. \ref{fig:PDFs} leads to a similar behaviour at $Q=3$ GeV, but this is not the case at $Q=5$ TeV. 
At high $Q$ values and low $x$, the {\sc NNPDF2.3QED}  prediction deviates from the other PDF sets. However this region is not probed in $t \bar t$ because at low $M$ the $gg$ luminosity is dominant w.r.t. the $g \gamma$ one. At large $x$ the {\sc NNPDF2.3QED} PDF set predicts a large photon PDF, while the {\sc CT14QED}, {\sc LUXqed} sets predict a small photon PDF.

\section{Calculation framework}
\label{sec:CalcFrame}

In this work we perform a NLO QCD+EW calculation for $t \bar t$ distributions using the {\sc MadGraph5\_aMC@NLO} framework \cite{Alwall:2014hca}. This is done by a simultaneous expansion in the two coupling constants ($\alpha_s, \alpha$). In eq. \ref{eq:Blobs} one can see a pictorial representation of this expansion including the new opening channels in each order as well as the definitions of the QCD, EW and QCD+EW observables.
\begin{equation}
\begin{aligned}[c]
&\text{LO} \;\;\;\;\;\;\;\;\;\;\;\; \overset{q\bar q, gg}{\underset{\rm LO~QCD}{\circled{$\alpha_s^2$}}} \;\;\;\;\;\;\; \overset{g\gamma}{\underset{\rm LO~EW}{\circled{$\alpha_s \alpha$}}} \;\;\;\;\;\;\; \overset{\gamma \gamma}{\circled{$\alpha^2$}}\\[1ex]
&\text{NLO}  \overset{qg}{\underset{\rm NLO~QCD}{\circled{$\alpha_s^3$}}} \;\;\;\;\; \overset{g\gamma,q\gamma}{\underset{\rm NLO~EW}{\circled{$\alpha_s^2 \alpha$}}} \;\;\;\;\; \circled{$\alpha_s \alpha^2$} \;\;\;\;\; \circled{$\alpha^3$}
\end{aligned}
\;\;\;\;\;\;\;\;
\begin{aligned}[c]
&\Sigma_{\rm QCD} \equiv \Sigma_{\rm  LO~QCD} + \Sigma_{\rm NLO~QCD} \;, \\[1ex]
&\Sigma_{\rm EW} \equiv \Sigma_{\rm LO~EW} + \Sigma_{\rm NLO~EW} \;, \\[1ex]
&\Sigma_{\rm QCD+EW} \equiv \Sigma_{\rm  QCD} + \Sigma_{\rm EW} \; .
\end{aligned}
\label{eq:Blobs}
\end{equation}
The LO EW and NLO EW orders are the leading ones in which the photon-induced contributions appear. It is checked that the effect of the subleading orders ($\alpha^2, \alpha_s \alpha^2, \alpha^3$) is below the percent level w.r.t. the $\alpha_s^2$ and they are neglected for this calculation.

The calculation is realised in the 5-flavour scheme and the EW parameters are defined in the $G_\mu$-scheme. The used input parameters are shown in eq. \ref{eq:param}.
\begin{align}
m_t &= 173.3 \text{ GeV}\, , \, m_H = 125.09 \text{ GeV} \, , \, m_W = 80.385 \text{ GeV} \, , \, m_Z = 91.1876 \text{ GeV} \,  , \nonumber \\
G_\mu &= 1.1663787 \cdot 10^{-5} \text{ GeV}^{-2} \, , \, \mu = \frac{H_T}{2} = \frac{1}{2} \sum_{i} m_{T,i} \; .
\label{eq:param}
\end{align}
In the definition of the renormalisation ($\mu_r$) and factorisation ($\mu_f$) scale, the possible extra jet or photon emission is included in the sum. The theoretical uncertainties are evaluated via an independent variation in the interval $\{\mu/2 < \mu_f,\mu_r <2\mu \}$. For the results presented in the following section we have chosen to use the {\sc NNPDF2.3QED} PDF set, where we also set the photon PDF artificially equal to zero, and we further compare with the results obtained with the {\sc CT14QED} PDF set.

\section{Differential distributions}

\begin{figure}[htb!]
\centering
\includegraphics[height=3.in]{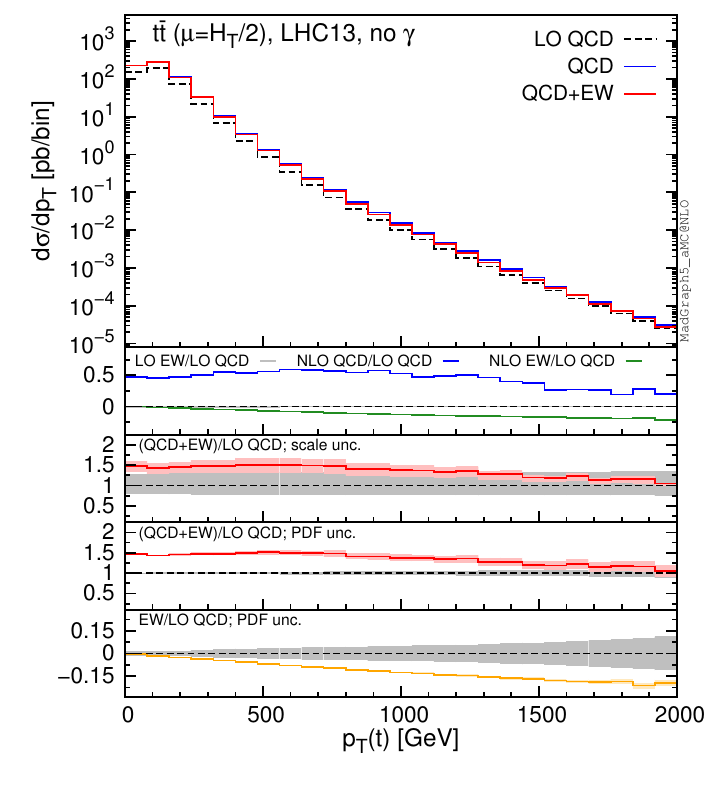}
\includegraphics[height=3.in]{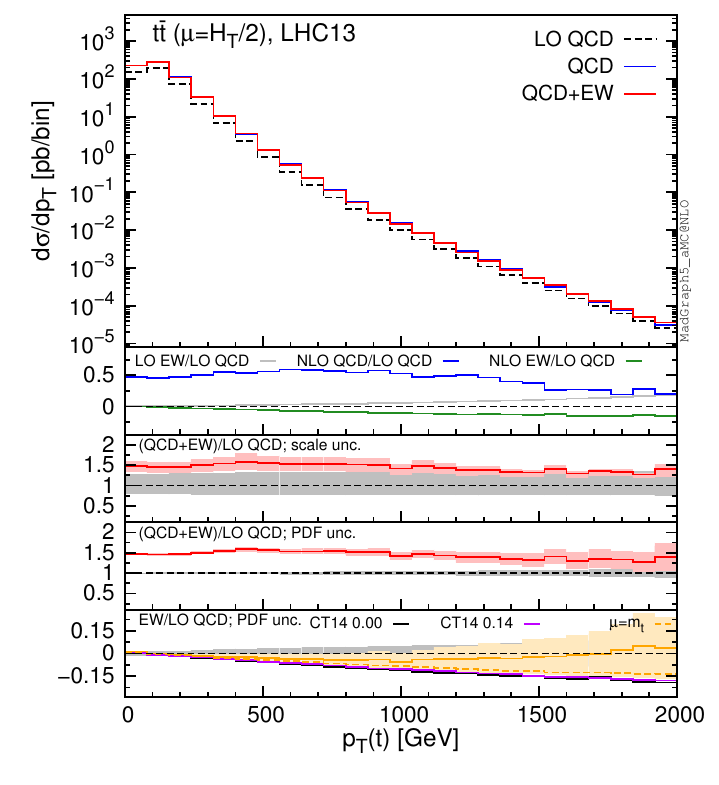}
\includegraphics[height=3.in]{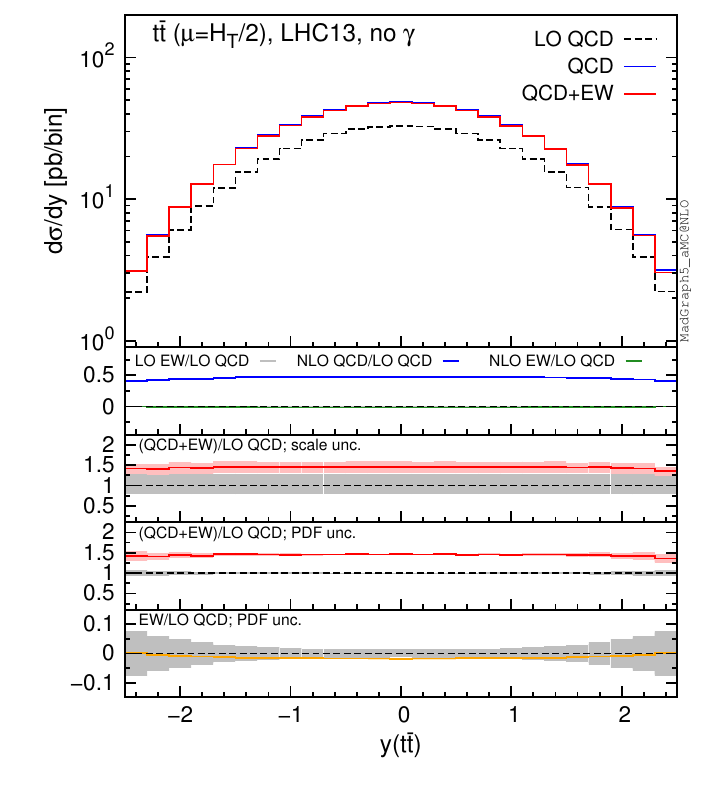}
\includegraphics[height=3.in]{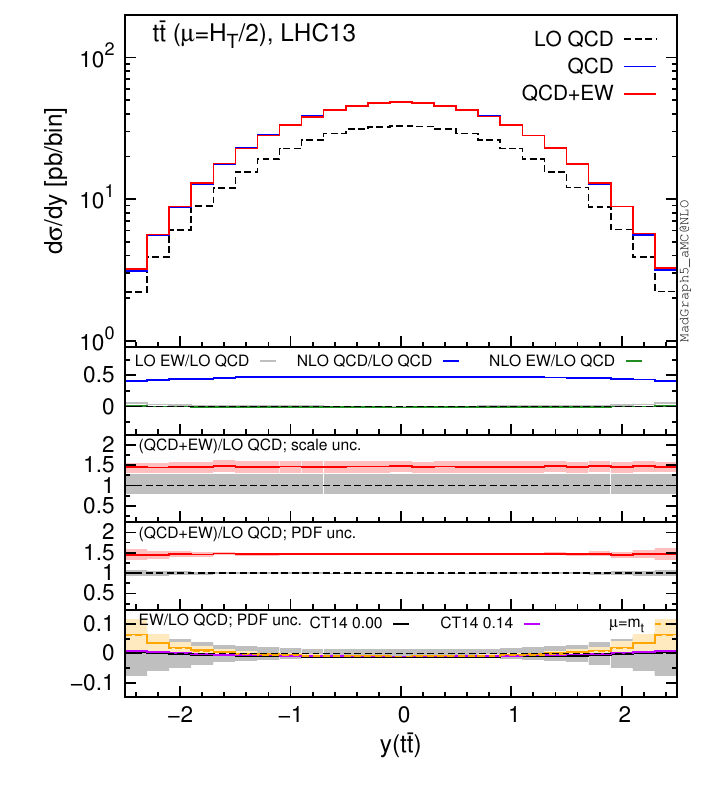}
\caption{$p_T(t)$ (top) and $y(t \bar t)$ (bottom) distributions at 13 TeV. The format of the plots is explained in the text.}
\label{fig:Distrib}
\end{figure}

From all the $t \bar t$ distributions at various energies (8, 13, 100 TeV) studied in \cite{Pagani:2016caq}, we restrict ourselves here to the transverse momentum of the top quark and the rapidity of the top-quark pair at 13 TeV, shown in fig. \ref{fig:Distrib}. The format of the plots is the following. In the main panel there are the distributions of the LO QCD (dashed black), the QCD (blue) and the QCD+EW (red). 
All the four insets are ratios w.r.t. the LO QCD. In the first inset there are the other three perturbative orders considered in section \ref{sec:CalcFrame}. The second and third inset show the QCD+EW accompanied with scale and PDF uncertainties respectively. The fourth inset shows only the EW accompanied with PDF uncertainties. The results are obtained using the {\sc NNPDF2.3QED} PDF set. On the left hand side of fig. \ref{fig:Distrib} the photon PDF is set artificially equal to zero whereas the right hand side includes the photon PDF. Finally in the last inset of the right hand side plots we compare the results with the ones obtained by using the {\sc CT14QED} PDF set.

Looking at the first inset of the $p_T(t)$ distributions and going from left to right we see that the photon PDF effect is mostly due to the LO EW contribution. The last inset of the left plot shows the Sudakov suppression mentioned in section \ref{sec:intro}. Moving to the right plot we see that for the {\sc NNPDF2.3QED} there is indeed a large cancelation between these Sudakov logarithms and the photon-induced contributions. However this is not the case for the {\sc CT14QED} PDF set, where the photon PDF impact is negligible. The last inset of the right $y(t \bar t)$ plot shows that in the {\sc NNPDF2.3QED} case there is a $\sim 5 \%$ effect of the photon PDF at large rapidity region. This region is already reached experimentally at 8 TeV \cite{Khachatryan:2015oqa,Aad:2015mbv} and it will be even more accurately measured at 13 TeV. On the other hand the {\sc CT14QED} prediction is also here equivalent with the {\sc NNPDF2.3QED} ``no photon'' scenario. Finally the two predictions are in agreement within the uncertainties, since in both cases the {\sc NNPDF2.3QED} prediction is accompanied with large uncertainties in the regions of interest. Similar differences from the comparison of these two PDF sets are pointed out also in \cite{Accomando:2016tah}, where the photon-initiated production of a di-lepton final state at the LHC is studied.

\section{Conclusions - Further research}

This work shows the effect of the EW corrections and especially of the photon-induced contributions on $t \bar t$ phenomenology, comparing the predictions of two different PDF sets. In the {\sc NNPDF2.3QED} PDF predictions, there is a large impact of the photon-induced contributions accompanied with large uncertainties. The {\sc CT14QED} PDF predictions, show a negligible impact of these contributions and lay to the lower limit of the {\sc NNPDF2.3QED} uncertainty band. At 13 TeV the photon-induced contributions in {\sc NNPDF2.3QED} are visible in rapidity but may be also in $p_T$ distributions, therefore the inclusion of NNLO QCD corrections is necessary in order to reduce the scale uncertainty of the prediction. The combination of the NNLO QCD and NLO EW calculations, using the more recent PDF sets {\sc NNPDF3.0QED} and {\sc LUXqed}, for $t \bar t $ distributions is in progress in collaboration with the co-authors of \cite{Pagani:2016caq} and the authors of \cite{Czakon:2016dgf}. 

\Acknowledgements

I am grateful to D. Pagani and M. Zaro for their collaboration on this work. My participation to this workshop was supported by the F.R.S.-FNRS ``Fonds de la Recherche Scientifique'' (Belgium).


\newpage
\begin{thebibliography}{99}

\bibitem{Pagani:2016caq} 
  D.~Pagani, I.~Tsinikos and M.~Zaro,
  Eur.\ Phys.\ J.\ C {\bf 76}, no. 9, 479 (2016)
  doi:10.1140/epjc/s10052-016-4318-z
  [arXiv:1606.01915 [hep-ph]].
  
\bibitem{Khachatryan:2015oqa} 
  V.~Khachatryan {\it et al.} [CMS Collaboration],
  Eur.\ Phys.\ J.\ C {\bf 75}, no. 11, 542 (2015)
  doi:10.1140/epjc/s10052-015-3709-x
  [arXiv:1505.04480 [hep-ex]].
  
\bibitem{Aad:2015mbv} 
  G.~Aad {\it et al.} [ATLAS Collaboration],
  Eur.\ Phys.\ J.\ C {\bf 76}, no. 10, 538 (2016)
  doi:10.1140/epjc/s10052-016-4366-4
  [arXiv:1511.04716 [hep-ex]].
  
\bibitem{Czakon:2013tha} 
  M.~Czakon, M.~L.~Mangano, A.~Mitov and J.~Rojo,
  JHEP {\bf 1307}, 167 (2013)
  doi:10.1007/JHEP07(2013)167
  [arXiv:1303.7215 [hep-ph]].
  
\bibitem{Czakon:2013goa} 
  M.~Czakon, P.~Fiedler and A.~Mitov,
  Phys.\ Rev.\ Lett.\  {\bf 110}, 252004 (2013)
  doi:10.1103/PhysRevLett.110.252004
  [arXiv:1303.6254 [hep-ph]].
  
\bibitem{Czakon:2015owf} 
  M.~Czakon, D.~Heymes and A.~Mitov,
  Phys.\ Rev.\ Lett.\  {\bf 116}, no. 8, 082003 (2016)
  doi:10.1103/PhysRevLett.116.082003
  [arXiv:1511.00549 [hep-ph]].
  
\bibitem{Czakon:2016dgf} 
  M.~Czakon, D.~Heymes and A.~Mitov,
  arXiv:1606.03350 [hep-ph].
  
\bibitem{Moch:2008qy} 
  S.~Moch and P.~Uwer,
  Phys.\ Rev.\ D {\bf 78}, 034003 (2008)
  doi:10.1103/PhysRevD.78.034003
  [arXiv:0804.1476 [hep-ph]].
  
\bibitem{Kidonakis:2009ev} 
  N.~Kidonakis,
  Phys.\ Rev.\ Lett.\  {\bf 102}, 232003 (2009)
  doi:10.1103/PhysRevLett.102.232003
  [arXiv:0903.2561 [hep-ph]].
  
\bibitem{Czakon:2009zw} 
  M.~Czakon, A.~Mitov and G.~F.~Sterman,
  Phys.\ Rev.\ D {\bf 80}, 074017 (2009)
  doi:10.1103/PhysRevD.80.074017
  [arXiv:0907.1790 [hep-ph]].
  
\bibitem{Ahrens:2010zv} 
  V.~Ahrens, A.~Ferroglia, M.~Neubert, B.~D.~Pecjak and L.~L.~Yang,
  JHEP {\bf 1009}, 097 (2010)
  doi:10.1007/JHEP09(2010)097
  [arXiv:1003.5827 [hep-ph]].
  
\bibitem{Beneke:2010da} 
  M.~Beneke, P.~Falgari and C.~Schwinn,
  Nucl.\ Phys.\ B {\bf 842}, 414 (2011)
  doi:10.1016/j.nuclphysb.2010.09.009
  [arXiv:1007.5414 [hep-ph]].
  
\bibitem{Kidonakis:2010dk} 
  N.~Kidonakis,
  Phys.\ Rev.\ D {\bf 82}, 114030 (2010)
  doi:10.1103/PhysRevD.82.114030
  [arXiv:1009.4935 [hep-ph]].
  
\bibitem{Ahrens:2011mw} 
  V.~Ahrens, A.~Ferroglia, M.~Neubert, B.~D.~Pecjak and L.~L.~Yang,
  JHEP {\bf 1109}, 070 (2011)
  doi:10.1007/JHEP09(2011)070
  [arXiv:1103.0550 [hep-ph]].
  
\bibitem{Ahrens:2011px} 
  V.~Ahrens, A.~Ferroglia, M.~Neubert, B.~D.~Pecjak and L.~L.~Yang,
  Phys.\ Lett.\ B {\bf 703}, 135 (2011)
  doi:10.1016/j.physletb.2011.07.058
  [arXiv:1105.5824 [hep-ph]].
  
\bibitem{Beneke:2011mq} 
  M.~Beneke, P.~Falgari, S.~Klein and C.~Schwinn,
  Nucl.\ Phys.\ B {\bf 855}, 695 (2012)
  doi:10.1016/j.nuclphysb.2011.10.021
  [arXiv:1109.1536 [hep-ph]].
  
\bibitem{Cacciari:2011hy} 
  M.~Cacciari, M.~Czakon, M.~Mangano, A.~Mitov and P.~Nason,
  Phys.\ Lett.\ B {\bf 710}, 612 (2012)
  doi:10.1016/j.physletb.2012.03.013
  [arXiv:1111.5869 [hep-ph]].
  
\bibitem{Pecjak:2016nee} 
  B.~D.~Pecjak, D.~J.~Scott, X.~Wang and L.~L.~Yang,
  Phys.\ Rev.\ Lett.\  {\bf 116}, no. 20, 202001 (2016)
  doi:10.1103/PhysRevLett.116.202001
  [arXiv:1601.07020 [hep-ph]].
  
\bibitem{Beenakker:1993yr} 
  W.~Beenakker, A.~Denner, W.~Hollik, R.~Mertig, T.~Sack and D.~Wackeroth,
  Nucl.\ Phys.\ B {\bf 411}, 343 (1994).
  doi:10.1016/0550-3213(94)90454-5
  
\bibitem{Kuhn:2006vh} 
  J.~H.~Kuhn, A.~Scharf and P.~Uwer,
  Eur.\ Phys.\ J.\ C {\bf 51}, 37 (2007)
  doi:10.1140/epjc/s10052-007-0275-x
  [hep-ph/0610335].
  
\bibitem{Kuhn:2013zoa} 
  J.~H.~Kuhn, A.~Scharf and P.~Uwer,
  Phys.\ Rev.\ D {\bf 91}, no. 1, 014020 (2015)
  doi:10.1103/PhysRevD.91.014020
  [arXiv:1305.5773 [hep-ph]].
  
\bibitem{Bernreuther:2005is} 
  W.~Bernreuther, M.~Fücker and Z.~G.~Si,
  Phys.\ Lett.\ B {\bf 633}, 54 (2006)
  Erratum: [Phys.\ Lett.\ B {\bf 644}, 386 (2007)]
  doi:10.1016/j.physletb.2006.11.052, 10.1016/j.physletb.2005.11.056
  [hep-ph/0508091].
  
\bibitem{Campbell:2016dks} 
  J.~M.~Campbell, D.~Wackeroth and J.~Zhou,
  [arXiv:1608.03356 [hep-ph]].
  
\bibitem{Hollik:2011ps} 
  W.~Hollik and D.~Pagani,
  Phys.\ Rev.\ D {\bf 84}, 093003 (2011)
  doi:10.1103/PhysRevD.84.093003
  [arXiv:1107.2606 [hep-ph]].
  
\bibitem{Kuhn:2011ri} 
  J.~H.~Kuhn and G.~Rodrigo,
  JHEP {\bf 1201}, 063 (2012)
  doi:10.1007/JHEP01(2012)063
  [arXiv:1109.6830 [hep-ph]].
  
\bibitem{Manohar:2012rs} 
  A.~V.~Manohar and M.~Trott,
  Phys.\ Lett.\ B {\bf 711}, 313 (2012)
  doi:10.1016/j.physletb.2012.04.013
  [arXiv:1201.3926 [hep-ph]].
  
\bibitem{Bernreuther:2012sx} 
  W.~Bernreuther and Z.~G.~Si,
  Phys.\ Rev.\ D {\bf 86}, 034026 (2012)
  doi:10.1103/PhysRevD.86.034026
  [arXiv:1205.6580 [hep-ph]].
  
\bibitem{Bernreuther:2010ny} 
  W.~Bernreuther and Z.~G.~Si,
  Nucl.\ Phys.\ B {\bf 837}, 90 (2010)
  doi:10.1016/j.nuclphysb.2010.05.001
  [arXiv:1003.3926 [hep-ph]].
  
\bibitem{Denner:2016jyo} 
  A.~Denner and M.~Pellen,
  JHEP {\bf 1608}, 155 (2016)
  doi:10.1007/JHEP08(2016)155
  [arXiv:1607.05571 [hep-ph]].
  
\bibitem{Hollik:2007sw} 
  W.~Hollik and M.~Kollar,
  Phys.\ Rev.\ D {\bf 77}, 014008 (2008)
  doi:10.1103/PhysRevD.77.014008
  [arXiv:0708.1697 [hep-ph]].
  
\bibitem{Martin:2004dh} 
  A.~D.~Martin, R.~G.~Roberts, W.~J.~Stirling and R.~S.~Thorne,
  Eur.\ Phys.\ J.\ C {\bf 39}, 155 (2005)
  doi:10.1140/epjc/s2004-02088-7
  [hep-ph/0411040].
  
\bibitem{Ball:2013hta} 
  R.~D.~Ball {\it et al.} [NNPDF Collaboration],
  Nucl.\ Phys.\ B {\bf 877}, 290 (2013)
  doi:10.1016/j.nuclphysb.2013.10.010
  [arXiv:1308.0598 [hep-ph]].
  
\bibitem{Bertone:2013vaa} 
  V.~Bertone, S.~Carrazza and J.~Rojo,
  Comput.\ Phys.\ Commun.\  {\bf 185}, 1647 (2014)
  doi:10.1016/j.cpc.2014.03.007
  [arXiv:1310.1394 [hep-ph]].
  
\bibitem{Bertone:2015lqa} 
  V.~Bertone, S.~Carrazza, D.~Pagani and M.~Zaro,
  JHEP {\bf 1511}, 194 (2015)
  doi:10.1007/JHEP11(2015)194
  [arXiv:1508.07002 [hep-ph]].
  
\bibitem{Schmidt:2015zda} 
  C.~Schmidt, J.~Pumplin, D.~Stump and C.~P.~Yuan,
  Phys.\ Rev.\ D {\bf 93}, no. 11, 114015 (2016)
  doi:10.1103/PhysRevD.93.114015
  [arXiv:1509.02905 [hep-ph]].
  
\bibitem{Ball:2014uwa} 
  R.~D.~Ball {\it et al.} [NNPDF Collaboration],
  JHEP {\bf 1504}, 040 (2015)
  doi:10.1007/JHEP04(2015)040
  [arXiv:1410.8849 [hep-ph]].
  
\bibitem{Manohar:2016nzj} 
  A.~Manohar, P.~Nason, G.~P.~Salam and G.~Zanderighi,
  arXiv:1607.04266 [hep-ph].
  
\bibitem{Alwall:2014hca} 
  J.~Alwall {\it et al.},
  JHEP {\bf 1407}, 079 (2014)
  doi:10.1007/JHEP07(2014)079
  [arXiv:1405.0301 [hep-ph]].
  
\bibitem{Accomando:2016tah} 
  E.~Accomando, J.~Fiaschi, F.~Hautmann, S.~Moretti and C.~H.~Shepherd-Themistocleous,
  arXiv:1606.06646 [hep-ph].
 

\end{thebibliography}
\end{document}